\documentclass[reprint,longbibliography,
 amsmath,amssymb,superscriptaddress]{revtex4-2}
\usepackage{graphicx}
\usepackage{dcolumn}
\usepackage{bm}
\usepackage[colorlinks=true, citecolor=blue, linkcolor=blue, urlcolor=blue]{hyperref}
\usepackage{xcolor}
\usepackage{amsmath, amssymb, amsfonts}
\usepackage{mathtools}
\usepackage{subfigure}
\usepackage{mathrsfs}
\usepackage{sidecap}
\usepackage{verbatim}
\newcommand{\col}[1]{\textcolor{red}{#1}}
\newcommand{\colb}[1]{\textcolor{blue}{#1}}
\definecolor{purple1}{rgb}{128,0,128}

\usepackage{bigints}
\newcommand{\bea}{\begin{eqnarray}}
\newcommand{\ea}{\end{eqnarray}}

\definecolor{darkpastelgreen}{rgb}{0.01, 0.75, 0.24}

\def\E{\mathbf{E}}
\def\B{\mathbf{B}}

\def\x{\mathbf{x}}

\def\d{\mathrm{d}}

\begin{document}

\title{
Macroscopic quantum superpositions in superconducting circuits} 
\author{Vitoria A. \surname{de Souza}}
\email{vitoriaa.fisica@gmail.com}
\affiliation{International Center of Physics, Institute of Physics, University of Brasilia, 70297-400 Brasilia, Federal District, Brazil} 
\author{Caio C. \surname{Holanda Ribeiro}}
\email{caiocesarribeiro@alumni.usp.br}
\affiliation{International Center of Physics, Institute of Physics, University of Brasilia, 70297-400 Brasilia, Federal District, Brazil} 
\author{Vitorio A. \surname{De Lorenci}}
\email{delorenci@unifei.edu.br}
\affiliation{Instituto de F\'{\i}sica e Qu\'{\i}mica, Universidade Federal de Itajub\'a, Itajub\'a, Minas Gerais 37500-903, Brasil}
%

\date\today

\begin{abstract}
A possible route to test whether macroscopic systems can acquire quantum features using superconducting circuits is here presented. It is shown that under general assumptions a classical test current pulse of fixed energy and adjustable length acquires quantum features after interacting with the quantum vacuum of the photon field. Further, it is shown that the mere existence of vacuum fluctuations can lead to the breakdown of energy and momentum conservation, and as the length of the pulse grows with respect to the characteristic size of the quantum system, the test pulse undergoes quantum-to-classical transition. This model differs from previous ones for its simplicity and points towards a new way of creating correlated systems suitable for quantum-based technology.
\end{abstract}

\maketitle

\emph{Introduction.}--- Ordinary experience with quantum mechanics suggests that quantum fluctuations in ``macroscopic systems'' are somewhat negligible, with Schr\"odinger's \cite{Schrodinger} famous thought experiment showcasing how strange can be the application of the quantum theory ideas to classical systems. Since Schr\"odinger's article, a cat state then designates a macroscopic system devised to test the limits of quantum mechanics, thus hinting towards ever more efficient quantum based-protocols and devices. The list of existing works on how to create cat states is vast \cite{vastArndt2014,vastClerk,vastDemid,vastHou2016,vastHuang,vastKorsbakken,vastReid,vastTeufel,vastZhou,penrose2003,Bassi,Caspar,Anastopoulos_2015,Donadi2021,RevModPhys,hauer,Feynman1957,Marletto2017,Ford}. We cite, for instance, the (theoretical) possibility of superposing mirrors in optomechanic systems \cite{penrose2003,Bassi}, the superposition of current states \cite{Caspar}, cat states in the context of gravity \cite{Anastopoulos_2015} and as probes of gravity induced-spatial decoherence \cite{Donadi2021}. See \cite{RevModPhys} for an extended review. Nevertheless, even though quantum technology and quantum metrology have already reached unprecedented goals, the emergence of the classical world from the ubiquitous quantum mechanics remains a mysterious phenomenon \cite{hauer}.

Of particular importance to the present work is the connection between gravity and cat states. We recollect one of R. Feynman's thought experiments to determine the gravitational field corresponding to a cat state \cite{Feynman1957,Marletto2017}. 
As Feynman reasoned in 1957, the only way to avoid quantizing gravity is to assume that quantum mechanics does not work for all scales. Since then, still no experiment was able to fully disclose the role played by gravity in quantum mechanical systems, if any. Another subtle aspect linked to the emergence of gravity in quantum systems is that of stress tensor fluctuations. It is a well-known fact that the energy content of a field, encapsulated in the energy momentum tensor $T_{\mu\nu}$, undergoes quantum fluctuations, and that should leave fingerprints on the spacetime geometry. For instance, light rays mimicking cat states can experience lightcone fluctuations \cite{Ford}. 

In this letter, an experiment inspired by the model of \cite{Ford} is proposed, in which superconducting circuits are used to probe the quantum features of objects of variable sizes. Schrödinger's original idea is further explored by enabling the observable being probed to have its length adjusted and compared to some external quantum system of reference. This allows for the assessment of the quantum-to-classical transition induced by the system size. In Schrödinger's thought experiment, the observable, which is the cat's health situation, pertains to an essentially macroscopic system. In our model, the system playing the role of the cat can, in principle, have controllable size, offering an interesting route with respect to previous works. For instance, a limitation in the experiment proposed in \cite{penrose2003}, where a macroscopic mirror was used as a probe of quantum fluctuations, is the impossibility of adjusting the mirror's size at will.

We propose to create and probe quantum fluctuations of macroscopic systems formed after some classical test current pulse travels along a transmission line affected by the vacuum fluctuations of an external electromagnetic field. We note that such vacuum fluctuations can be harnessed, for instance, to induce superconductivity \cite{cavity} and the Casimir effect \cite{kimball}. Therefore, by adjusting the classical properties of the incident pulse, one can effectively determine how quantum features of the external system are transferred to the incident pulse.   

\begin{figure}[h!]
\center
\includegraphics[width=0.5\textwidth]{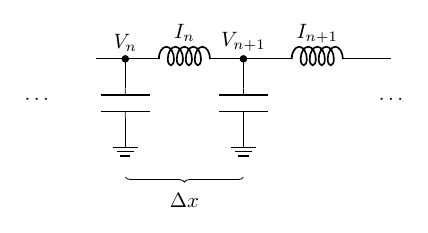}
\caption{The $n$-th cell of the electric circuit, comprised of a capacitor and an inductor, and all the cells are identical, with capacitance and inductance equal to $C_0$ and $L_0$, respectively. $I_n$ and $V_n$ are the electric current and potential in the cell as indicated. The full circuit is assumed to be sufficiently long, in such a way that the number of cells is formally infinite. Furthermore, magnetic coupling between different inductors is assumed negligible. Here, $\Delta x$ is the cell-to-cell distance. 
}
\label{fig1}
\end{figure}

\emph{The model.}--- Within the realm of classical electromagnetism, we consider the periodic structure formed by the repetition of the cell depicted in Fig.~\ref{fig1},
where $I_{n}$ and $V_n$ are the electric current and potential, respectively, of the $n$-th cell, as indicated. We refer to \cite{Blencowe} for a recent application of similar circuits to analogue gravity. By working with the (electric charge) variable $\psi_n$ such that $I_n=\d\psi_n/\d t$, current conservation leads to $V_{n+1}=-(1/C_{0})(\psi_{n+1}-\psi_n)$, whereas Faraday's law leads to the Lagrangian for $\psi_n$
%
%
%
%
%
\begin{align}
L=\sum_{n}\bigg[\frac{L_0}{2}\left(\frac{\d\psi_n}{\d t}\right)^2-\frac{1}{2C_0}(\psi_{n+1}-\psi_{n})^2-\psi_n\frac{\d \Phi_{n}}{\d t}\bigg],
\end{align}
where $L_0$ and $C_0$ are the (constant) self-induction and capacitance of the cells, and $\Phi_{n}$ is the magnetic flux caused by an applied external field.  
We are interested in studying long wavelength collective excitations in this circuit, i.e., whose characteristic wavelength $\lambda$ is much larger than the cell-to-cell distance $\Delta x$. Accordingly, by placing each cell at the coordinate $x_n\equiv n \Delta x$, setting $L_0=\Delta x\gamma_L$, $C_0=\Delta x\gamma_C$, and letting the functions $f=f(t,x)$ and $\psi=\psi(t,x)$ be such that $\psi_n(t)=\sqrt{\gamma_C}\psi(t,x_n)$, $\d\Phi_n(t)/\d t=\Delta xf(t,x_n)/\sqrt{\gamma_C}$, the limit $\Delta x\rightarrow0$ implies $L\rightarrow \int\d x\mathcal{L}$, where
\begin{align}
\mathcal{L}=\frac{1}{2c^2}\left(\frac{\partial\psi}{\partial t}\right)^2-\frac{1}{2}\left(\frac{\partial\psi}{\partial x}\right)^2-\psi f,\label{lagrangian}
\end{align}
and $c=1/\sqrt{\gamma_L\gamma_C}$ has units of velocity. Note that $\gamma_L$ and $\gamma_C$ are constants interpreted as inductance and capacitance densities, respectively. We assume in what follows that units are such that $c=\hbar=1$, and thus the Euler-Lagrange equation becomes
\begin{equation}
\partial^2_t\psi-\partial^2_x\psi=- f,\label{fieldeqnew}
\end{equation}
which is the basic field equation for the model. 

Our reasoning is based on the energy and momentum in this system. By defining the two vector $x^{\mu}=(t,x)$, $\mu=0,1$, and the metric $\eta_{\mu\nu}=\mbox{diag}(1,-1)$, the conservation laws are obtained from the canonical energy-momentum tensor \cite{Maggiore}
\begin{equation}
    T^{\mu}_{\ \nu}=\frac{\partial\mathcal{L}}{\partial(\partial_\mu\psi)}\partial_\nu\psi-\delta^{\mu}_{\ \nu}\mathcal{L},
\end{equation}
which, by means of the Euler-Lagrange equation, satisfies
\begin{equation}
    \partial_\mu T^{\mu}_{\ \nu}=\psi\partial_\nu f.\label{continuity}
\end{equation}
The force density $\psi\partial_\nu f$ therefore plays the same role as $F_{\mu\nu}J^{\nu}$ in electrodynamics \cite{wald2022}, and measures the energy-momentum exchange rate between the circuit and the external source.

The total momentum 2-vector $P^{\mu}$ is then defined by $P^{\mu}=\int\d x T^{0\mu}$, and it follows from Eq.~\eqref{continuity} that $P^{0}$ and $P^{1}$ are conserved for {\it all} processes whenever $\partial_\nu f=0$. 
%
%
Thus 
\begin{equation}
H=P^{0}=\frac{1}{2}\int\d x\left[(\partial_t\psi)^2+(\partial_x\psi)^2\right],\label{energy}
\end{equation}
is the self-energy of the system, whereas
\begin{equation}
    P=P^1=-\int\d x (\partial_t\psi)(\partial_x\psi)\label{momentum}
\end{equation} is its momentum.

\emph{Langevin equation.}--- We let the circuit portion inside the region $-\ell/2<x<\ell/2$ be influenced by an external magnetic field, which produces a current through the source term $f$ in Eq.~\eqref{fieldeqnew}. Accordingly, the general solution of the field equation assumes the form
\begin{equation}
    \psi(t,x)=\psi_{\rm c}(t,x)+\psi_{\rm q}(t,x),\label{totalsol}
\end{equation}
where $\psi_{\rm c}$ is such that $(\partial^2_t-\partial^2_x)\psi_{\rm c}=0$ and models test pulses in the circuit, whereas $\psi_{\rm q}$ is the particular solution produced by $f$ and it encompasses any fluctuations in the system whose origin can be traced back to $f$. $\psi_{\rm q}$ can be readily found by means of the (causal) propagator, $G(t,x;t',x')$, solution of
\begin{equation}
    (\partial_t^2-\partial_x^2)G(t,x;t',x')=-\delta(t-t')\delta(x-x'),
\end{equation}
and given by \cite{wald2022}
\begin{equation}
    G(t,x;t',x')=\frac{1}{(2\pi)^2}\int_{-\infty}^\infty\d\omega\int_{-\infty}^\infty\d k\frac{e^{-i\omega\Delta t+i k\Delta x}}{(\omega+i\epsilon)^2-k^2}.\label{prop}
\end{equation}
Here, $\epsilon>0$ is a small number to be taken to zero later on, at the end of the calculations. Thus,
\begin{equation}
    \psi_{\rm q}(t,x)=\int\d^2x'G(t,x;t',x')f(t',x').\label{langevinsol}
\end{equation}

We assume that the a monochromatic electromagnetic wave of frequency $\omega_{\rm e}$ is sent towards the circuit in the region $-\ell/2<x<\ell/2$  and whose magnetic field is given (on the circuit) by
\begin{equation}
       \B(t,\x)=\B_{\omega_{\rm e}}ae^{-i\omega_{\rm e}t}+\B^*_{\omega_{\rm e}}a^{*}e^{i\omega_{\rm e}t}.\label{externalmode}
\end{equation}
Furthermore, we assume that the effect of this external field on the circuit occurs only at the fundamental level of Faraday's law, i.e., it contributes with a magnetic flux $\Phi_n$ to the $n$-th cell. It should be stressed however that, qualitatively, the same results can be obtained if other types of interactions are also present. For instance, the electric properties of the circuit can be taken to depend on the external wave via the archetypal Rabi-like interaction \cite{Braumuller2017}.  

With the field \eqref{externalmode}, it is straightforward to obtain the source term
\begin{equation}
    f=-i\omega_{\rm e}\sqrt{\gamma_C}\left(\phi ae^{-i\omega_{\rm e}t}-\phi^*a^{*}e^{i\omega_{\rm e}t}\right),
\end{equation}
where $\phi$ is the (constant) magnetic flux in each inductor. Therefore, Eq.~\eqref{prop} can be used to find
\begin{equation}
    \psi_{\rm q}(t,x)=-\sqrt{\gamma_C}\frac{\sin(\omega_{\rm e}\ell/2)}{\omega_{\rm e}}\left[\phi ae^{-i\omega_{\rm e}(t-x)}+c.c.\right],\label{partsol}
\end{equation}
valid for $x>\ell/2$. We now let the external magnetic field to be a quantum field by letting $a$ become an operator subjected to $[\hat{a},\hat{a}^\dagger]=1$. The theory vacuum state, $|0\rangle$, is defined by $\hat{a}|0\rangle=0$, and it is such that $\langle \hat{f}\rangle=0$ and $\langle \hat{f}(t,x)\hat{f}(t',x')\rangle \neq0$. In this case, the field equation Eq.~\eqref{fieldeqnew} becomes a Langevin equation, whose solution is given by Eq.~\eqref{langevinsol}. Therefore, we find that $\hat{\psi}_{\rm q}$ and all quantities built from it also undergo quantum fluctuations, with $\langle\hat{\psi}_{\rm q}\rangle=0$ and $\langle \hat{\psi}_{\rm q}(t,x)\hat{\psi}_{\rm q}(t',x')\rangle \neq0$.

\emph{Energy-momentum tensor fluctuations.}--- We are interested in studying the effect of the external field upon a current pulse sent from $x\rightarrow -\infty$ towards a measuring device at $x\rightarrow \infty$ that probes the energy and momentum in the system. Thus, the quantities of interest are $\hat{H}$ and $\hat{P}$ for $t\rightarrow\infty$. By plugging the solution \eqref{totalsol} back into Eqs.~\eqref{energy} and \eqref{momentum}, we find that $\hat{H}=H_{\rm c}+\hat{H}_{\rm q}+\hat{H}_{\rm m}$ and $\hat{P}=P_{\rm c}+\hat{P}_{\rm q}+\hat{P}_{\rm m}$, where $H_{\rm c}$, $P_{\rm c}$ and $\hat{H}_{\rm q}$, $\hat{P}_{\rm q}$ are the energy and momentum in the system in the absence of $\hat{\psi}_{\rm q}$ and $\psi_{\rm c}$, respectively, and  
\begin{align}
    \hat{H}_{\rm m}&=\int\d x[(\partial_t\psi_{\rm c})(\partial_t\hat{\psi}_{\rm q})+(\partial_x\psi_{\rm c})(\partial_x\hat{\psi}_{\rm q})],\\
    \hat{P}_{\rm m}&=-\int\d x[(\partial_t\psi_{\rm c})(\partial_x\hat{\psi}_{\rm q})+(\partial_x\psi_{\rm c})(\partial_t\hat{\psi}_{\rm q})],
\end{align}
are mixed contributions.  In particular, $\hat{H}$ and $\hat{P}$ are Hermitian operators whose fluctuations can be probed. 

Because the background field is prepared in a vacuum state, we find on average that $\langle \hat{H}\rangle=H_{\rm c}+\langle \hat{H}_{\rm q}\rangle$ and $\langle \hat{P}\rangle=P_{\rm c}+\langle \hat{P}_{\rm q}\rangle$, and thus the background field contributes with residual energy and momentum to the system in the absence of classical currents.  Moreover, if $\Delta \mathcal{O}$ is the uncertainty in the observable $\hat{\mathcal{O}}$, we find
\begin{align}
(\Delta H)^2&=(\Delta H_{\rm q})^2+\langle \hat{H}_{\rm m}^2\rangle,\\
(\Delta P)^2&=(\Delta P_{\rm q})^2+\langle \hat{P}_{\rm m}^2\rangle.
\end{align}
Notice the presence of the mixed contributions in the squared uncertainties, which accounts for the interaction between the classical and the quantum part of the current. In particular, if the background field is kept fixed, the only degree of freedom in changing the fluctuations is the classical pulse, by means of $\hat{H}_{\rm m}$ and $\hat{P}_{\rm m}$. We adopt the classical current-induced shift on top of the ever present background noise in the squared uncertainties of $\hat{H}$ and $\hat{P}$ as a probe of the degree of ``quantumness'' in the transmitted current pulse. 

In order to illustrate a sufficiently interesting scenario, let us consider at $t\rightarrow-\infty$ an incident Gaussian wave packet of characteristic length $\sigma$ 
\begin{equation}
\psi_{\rm c}(t,x)=\sqrt{\frac{2\sigma E_0}{\sqrt{\pi}}}e^{-(x-t)^2/(2\sigma^2)}.\label{gaussianpacket}
\end{equation}
%
The energy and momentum of the incident pulse are found from Eqs.~\eqref{energy} and \eqref{momentum} to be
\begin{equation}
H_{\rm c}=P_{\rm c}=E_0,
\end{equation}
and this, together with $\sigma$, are the only adjustable experimental parameters of the classical pulse.

By gathering all expressions, $\langle \hat{H}_{\rm m}^2\rangle$ and $\langle \hat{P}_{\rm m}^2\rangle$ can be calculated exactly. We find that, in dimensionful units,
\begin{equation}
   \langle \hat{P}_{\rm m}^2\rangle= \langle \hat{H}_{\rm m}^2\rangle=\hbar\omega_{\rm e}E_0\alpha \left(\frac{\omega_{\rm e}\sigma}{c}\right)^3e^{-(\omega_{\rm e}\sigma/c)^2},\label{result}
\end{equation}
where $\alpha$ is a dimensionless constant depending only on the circuit properties and external field parameters,
\begin{equation}
    \alpha=4\sqrt{\pi}\left[\frac{\sin \omega_{\rm e}\ell/(2c)}{\omega_{\rm e}\ell/(2c)}\right]^2\frac{c\ell^2\gamma_C|\phi|^2}{\hbar}.
\end{equation}
%
%

Equation \eqref{result} is the major result of our model. Notice that the uncertainty in probing the energy of the transmitted pulse is maximum for $\omega_{\rm e}\sigma=c\sqrt{3/2}$, i.e., for incident pulses whose length $\sigma$ is comparable to the size of ``quantum current length'' $\sigma_e\equiv c/\omega_{\rm e}$. Furthermore, as $\sigma\gg\sigma_e$, the uncertainty in measuring energy and momentum is suppressed, indicating that as the current pulse becomes macroscopic with respect to the quantum system it can no longer carry information about the wave function of the incident photon. This supports the view that superposition of classical objects like spacetime geometries in quantum gravity can only be probed in distance scales comparable to the size of the quantum system generating the superposition \cite{Feynman1957}. We also call attention to the effect of the classical pulse energy $E_0$ in Eq.~\eqref{result}: quantum fluctuations are amplified by the presence of the classical signal. This suggests the possibility of using classical physics to enhance quantum correlations for quantum-based devices. A similar effect was explored in the context of quantum optics in \cite{ford2017}.   

We now turn our attention to a second aspect revealed by Eq.~\eqref{result}. In the absence of quantum fluctuations of the background field, this scenario is characterized by complete, non-fluctuating transmission of emitted signals. When the background field is in its quantum vacuum state, however, on average only classical pulses are detected in the circuit, but the mere superposition of the classical pulse with the vacuum fluctuations leads to a fluctuation of both the energy of the transmitted pulse as well as the spontaneous breakdown of total energy and momentum conservation. Indeed, this breakdown is expected, since on average $\partial_\mu\langle T^{\mu}_{\ \nu}\rangle\neq0$ by means of Eq.~\eqref{continuity}. The interesting novel feature is the origin of the breakdown: the existence of fundamental vacuum fluctuations.

\emph{Final remarks.}---  We presented a possible route to create macroscopic quantum superposition of systems of adjustable sizes using superconducting circuits. We showed that test current pulses undergo quantum fluctuations after interacting with the quantum vacuum of an external electromagnetic field, and the degree of ``quantumness'' goes to zero if the characteristic sizes of the pulses increase in comparison to the quantum system characteristic size. Our model's advantages over previous proposals include the simplified protocol of correlating current pulses, and the source of the quantum fluctuations being the quantum vacuum, e.g., of the field in a cavity. We stress that our arguments can be easily adapted to different types of interactions between the electromagnetic and the circuit.


%
%

%
%

\emph{Acknowledgements.}--- 
C.C.H.R. would like to thank the {\it Funda\c{c}\~ao de Apoio \`a Pesquisa do Distrito
Federal} under grant 00193-00002051/2023-14. V.A.D.L. is supported in part by {\it Conselho Nacional de Desenvolvimento Cient\'{\i}fico e Tecnol\'ogico} under grant  302492/2022-4.

\bibliography{QGAv3}

\end{document}